\documentclass[a4paper]{jpconf}
\usepackage{graphicx}
\usepackage{citesort}
\usepackage{amssymb}

\begin{document}
\title{Physics Goals and Status of JEM-EUSO and its Test Experiments}

\author{Andreas Haungs$^1$ for the JEM-EUSO Collaboration$^2$}

\address{$^1$Institut f\"ur Kernphysik, Karlsruher Institut f\"ur Technologie (KIT), Germany}
\address{$^2$see http://jemeuso.riken.jp}

\ead{haungs@kit.edu}

\begin{abstract}
The JEM-EUSO mission aims to explore the origin of the extreme energy cosmic rays
(EECRs) through the observation of air-shower fluorescence light from space. The superwide-
field telescope looks down from the International Space Station onto the night sky to
detect UV photons (fluorescence and Cherenkov photons) emitted from air showers. Such a
space detector offers the remarkable opportunity to observe a huge volume of atmosphere at
once and will achieve an unprecedented statistics within a few years of operation. Several test
experiments are currently in operation: e.g., one to observe the fluorescence background from the
edge of the Atmosphere (EUSO-Balloon), or another to demonstrate on ground the
capability of detecting air showers with a EUSO-type telescope (EUSO-TA). In this contribution a
short review on the scientific objectives of the mission and an update of the instrument definition,
performances and status, as well as status of the test experiments will be given.
\end{abstract}

\section{Introduction}
\begin{figure}[ht]
\centering
\includegraphics[width=0.81\linewidth]{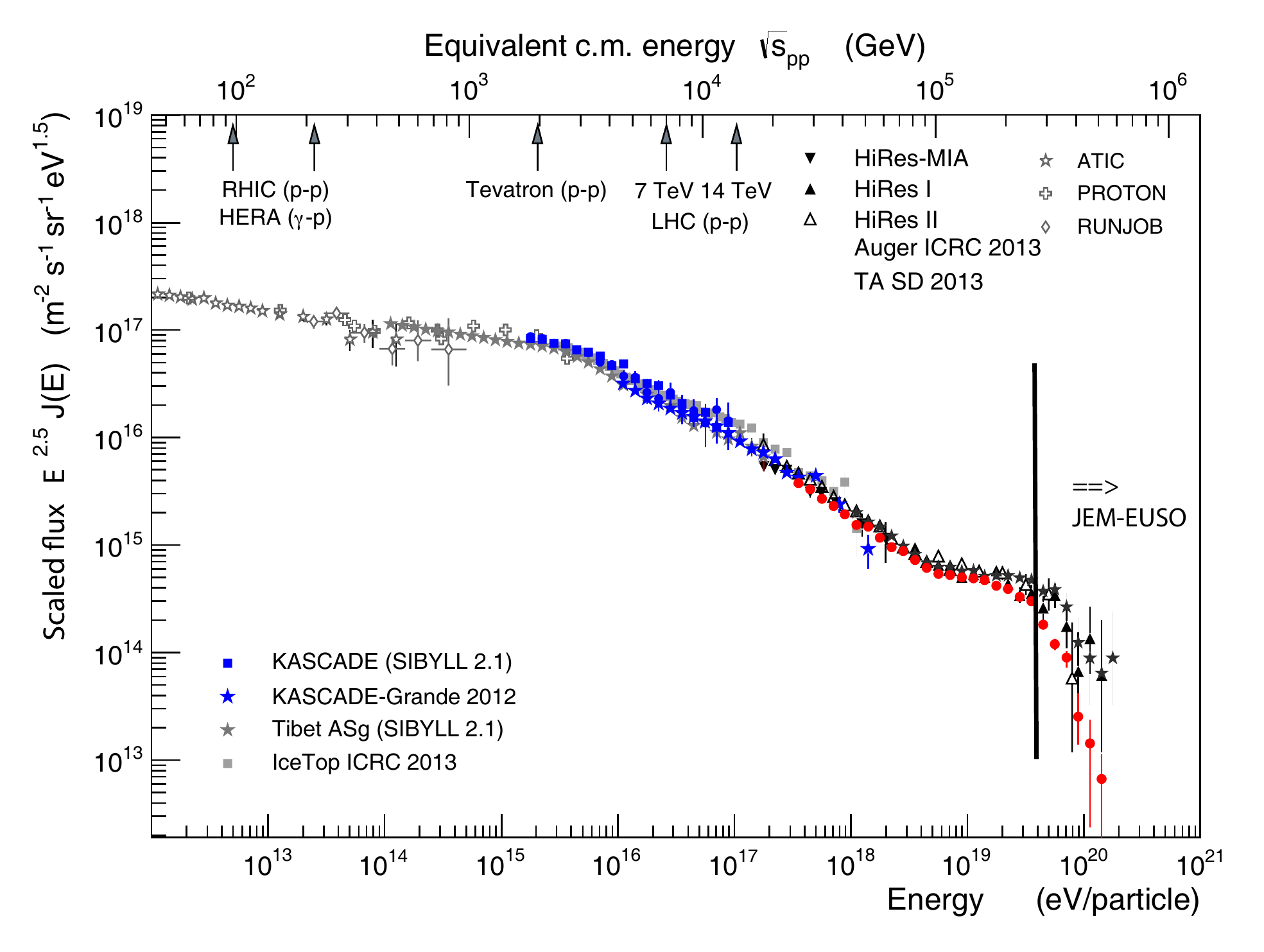}
\caption{The all-particle spectrum of cosmic rays as measured by various experiments. 
Largest uncertainties  are in the region of the highest energies: the target of the JEM-EUSO 
mission.}  
\label{fig_spec}
\end{figure}

The region of cosmic rays of the highest energies, i.e.~the Extreme Energy Cosmic Rays - EECR, 
of the wide spanning cosmic ray energy spectrum is the one least explored. 
This is natural as the flux of such particles reaching Earth is extremely low. 
In addition to the low statistical accuracy, the existing measurements disagree in their flux by 
a surprisingly large amount (see Fig.~\ref{fig_spec}).

The origin of the differences is still not fully understood and several causes may play a role.
One reason could lie in the fact that the energy reconstruction is limited by our 
current theoretical understanding of extensive air showers (EAS)~\cite{enterria}. 
These showers are generated by interaction of the impinging cosmic particle with the Atmosphere. 
The products of this are measured with detectors of large exposure and the data is interpreted in terms of  
the arrival direction, energy and mass of the primary particle.
The features of the first interaction (cross-sections and multiplicity distributions of secondaries)
as well as partly the kinematic features (extreme forward direction) of interactions 
at all energies relevant for the shower development are not well measured by accelerators. 
Hence, extrapolations are used in simulating the EAS leading to different interpretations 
if different hadronic interaction models or different observables are used.
Another cause could be that the two main experiments at the highest energies 
(Pierre Auger Observatory~\cite{pao} and Telescope Array~\cite{ta}) are located at different 
hemispheres and therefore, look at different regions of the sky. Several theories suggest that the EECR stem 
from astronomically nearby ($<100\,$Mpc) Active Galactic Nuclei, where it is well known that 
the distributions of those AGNs are not uniform over the sky~\cite{agn}.

For future experiments, large efforts have to be made to reach significantly higher statistics in 
measurements as well as a coverage of the whole sky. Both require a much higher exposure than 
that of existing experiments. 

The Extreme Universe Space Observatory (EUSO)
(Fig.~\ref{fig_principle}) at the Japanese Module (JEM) of the International Space Station (ISS) 
is the first space mission devoted to the scientific research of EECR~\cite{toshi,taka,marco,EAinstr}.

\section{JEM-EUSO}

\subsection{Scientific Goals}

\begin{figure}[ht]
\centering
\includegraphics[width=0.40\linewidth]{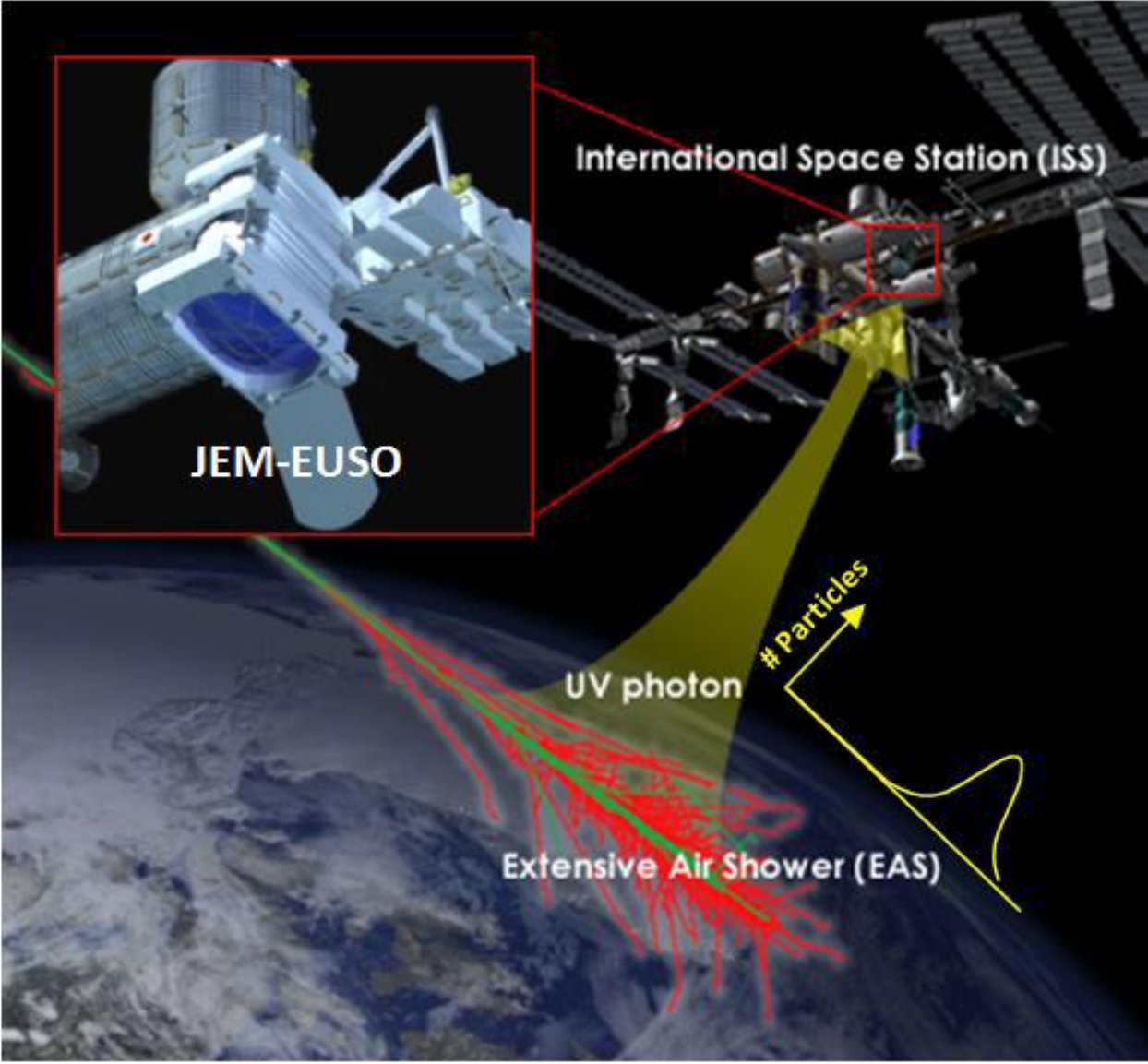}
\includegraphics[width=0.53\linewidth]{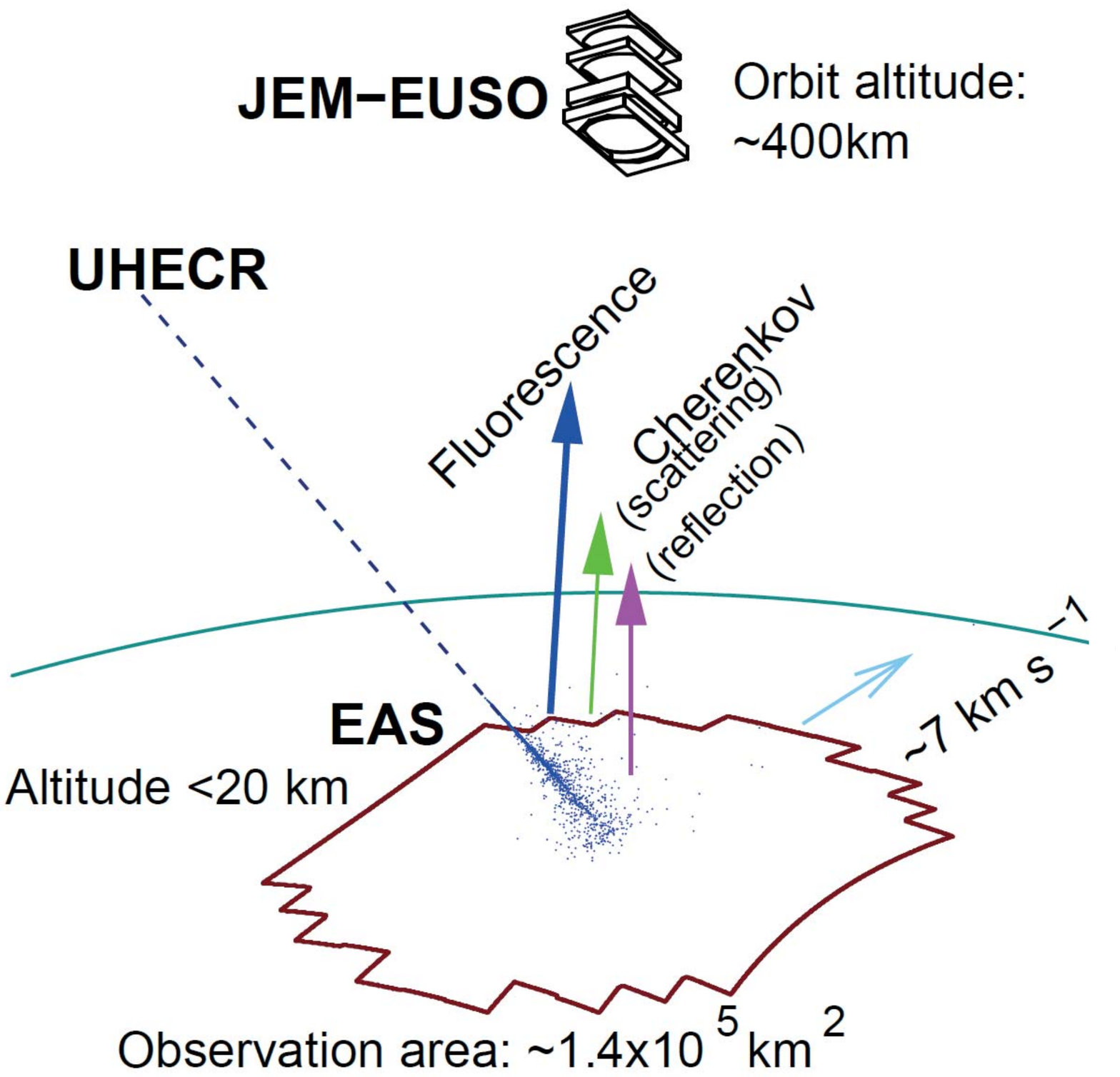}
\caption{Left: Principle of the JEM-EUSO mission to detect extreme energy
cosmic rays via the fluorescence emission of extensive air showers in the Atmosphere. 
Right: Schematic view of the JEM-EUSO mission: EASs generate fluorescence light that is emitted 
isotropically as well as Cherenkov light emitted in forward direction. 
The latter is also scattered and reflected and can reach the aperture of the telescope.} 
\label{fig_principle}
\end{figure}

Main goal is the exploration of the Universe through the detection of the extreme
energy cosmic rays and neutrinos by looking downward from the ISS to detect the fluorescence 
light of extensive air-showers that they generate in the Earth's atmosphere (Fig.~\ref{fig_principle}). 

The main scientific objective is astronomy and astrophysics through the particle channel. 
This requires to identify the sources of cosmic rays by the reconstruction of the arrival direction
and energy spectra with a high collecting power, beyond any other previous or
planned experiment so far. In addition, exploratory objectives were defined~\cite{perf-paper}.
This leads to the following physics program of JEM-EUSO:
\begin{enumerate}
\item astronomy and astrophysics through the particle channel at $E > 5 \times 10^{19}\,$eV;
\item the detection of extreme energy gamma rays;
\item the detection of extreme energy neutrinos; 
\item exploratory studies of the galactic magnetic fields;
\item fundamental physics studies (e.g. Lorentz invariance tests) at extreme energies;
\item global survey of nightglows, plasma discharges, lightning, meteors and other terrestrial
transient sources in UV light.
\end{enumerate}
The last one is possible due to the fast UV-light monitoring of the entire Atmosphere, 
which was never performed before.

Detailed simulations have shown that with three years of operation, JEM-EUSO will achieve a statistics 
of 100 events above $10^{20}\,$eV and that with these events, if exists, nearby sources 
(even with a heavy dominant elemental composition at those energies) will be identified~\cite{parizot}.

\subsection{Instrument}  

The JEM-EUSO instrument~\cite{EAinstr,kajino} consists of the telescope, the focal surface,  
a monitoring system for the atmospheric conditions, and a calibration system. 
In addition, there will be support and calibration systems on 
ground as well as at the ISS. 
\begin{figure}[t]
\centering
\includegraphics[width=0.40\linewidth]{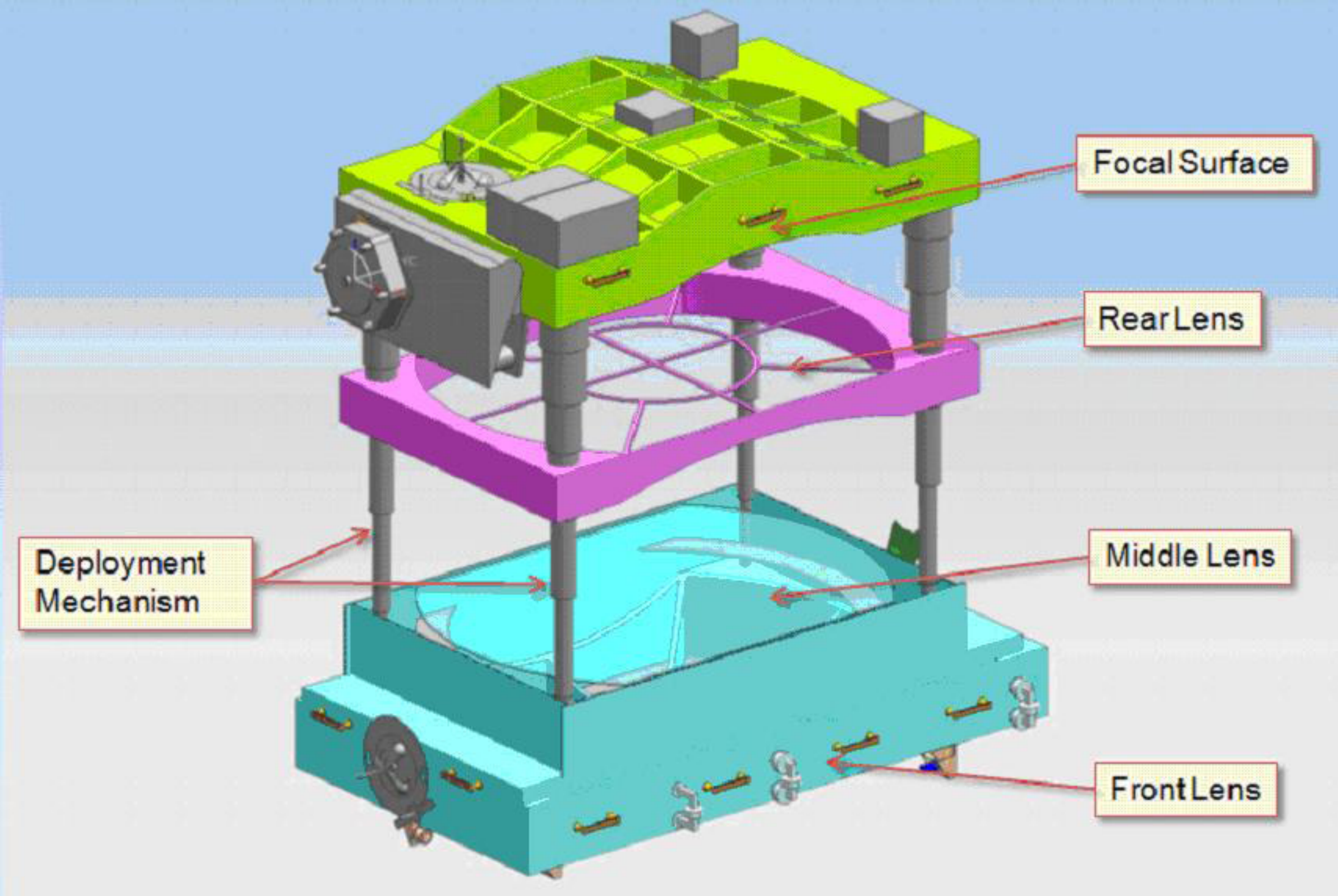}
\includegraphics[width=0.46\linewidth]{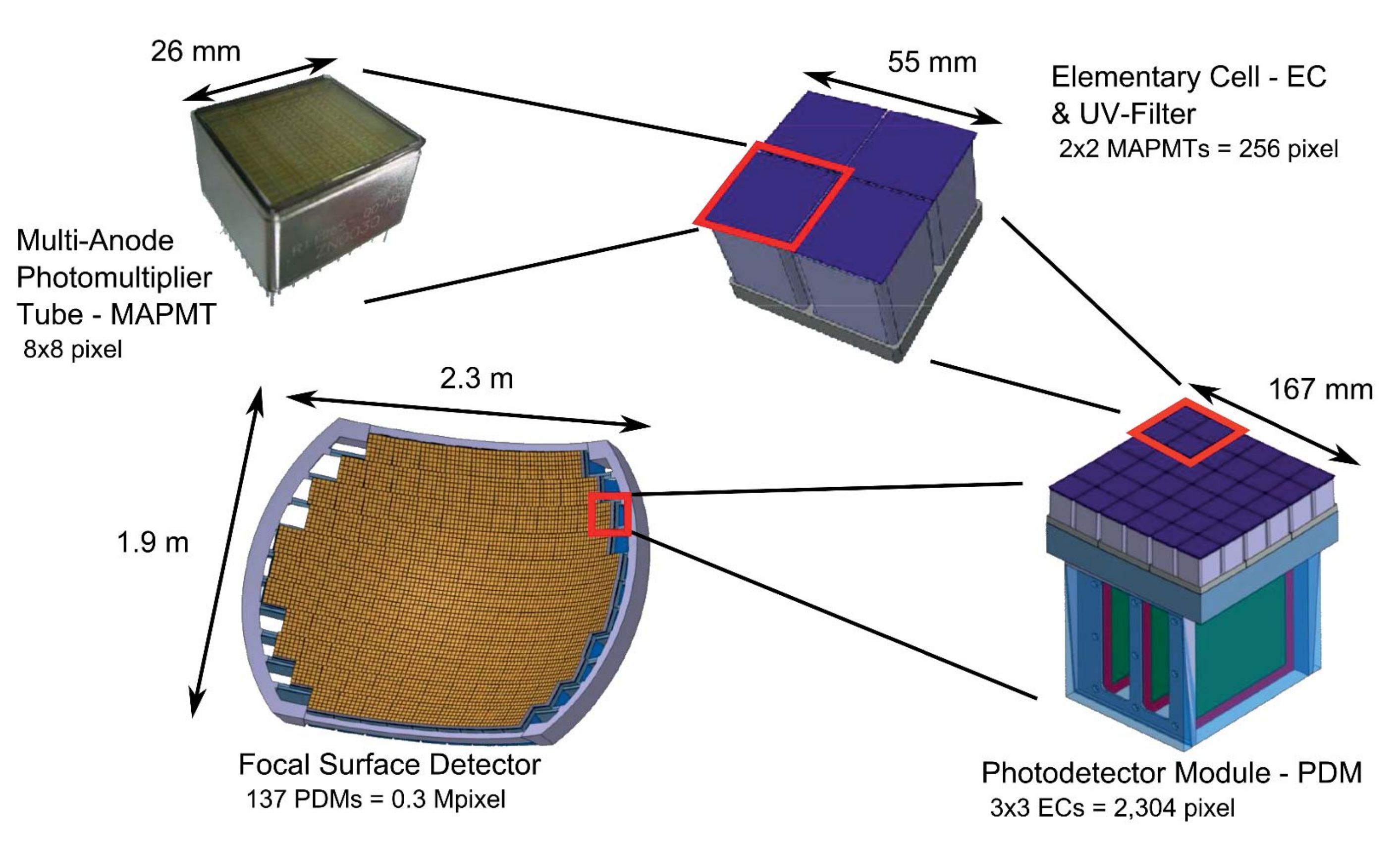}
\caption{Conceptual view of the JEM-EUSO telescope with its three lenses (left panel). 
Schematic view of the focal surface and it's components of the JEM-EUSO mission (right panel).} 
\label{fig_fs}
\end{figure}
\begin{figure}[ht]
\centering
\includegraphics[width=0.45\textwidth]{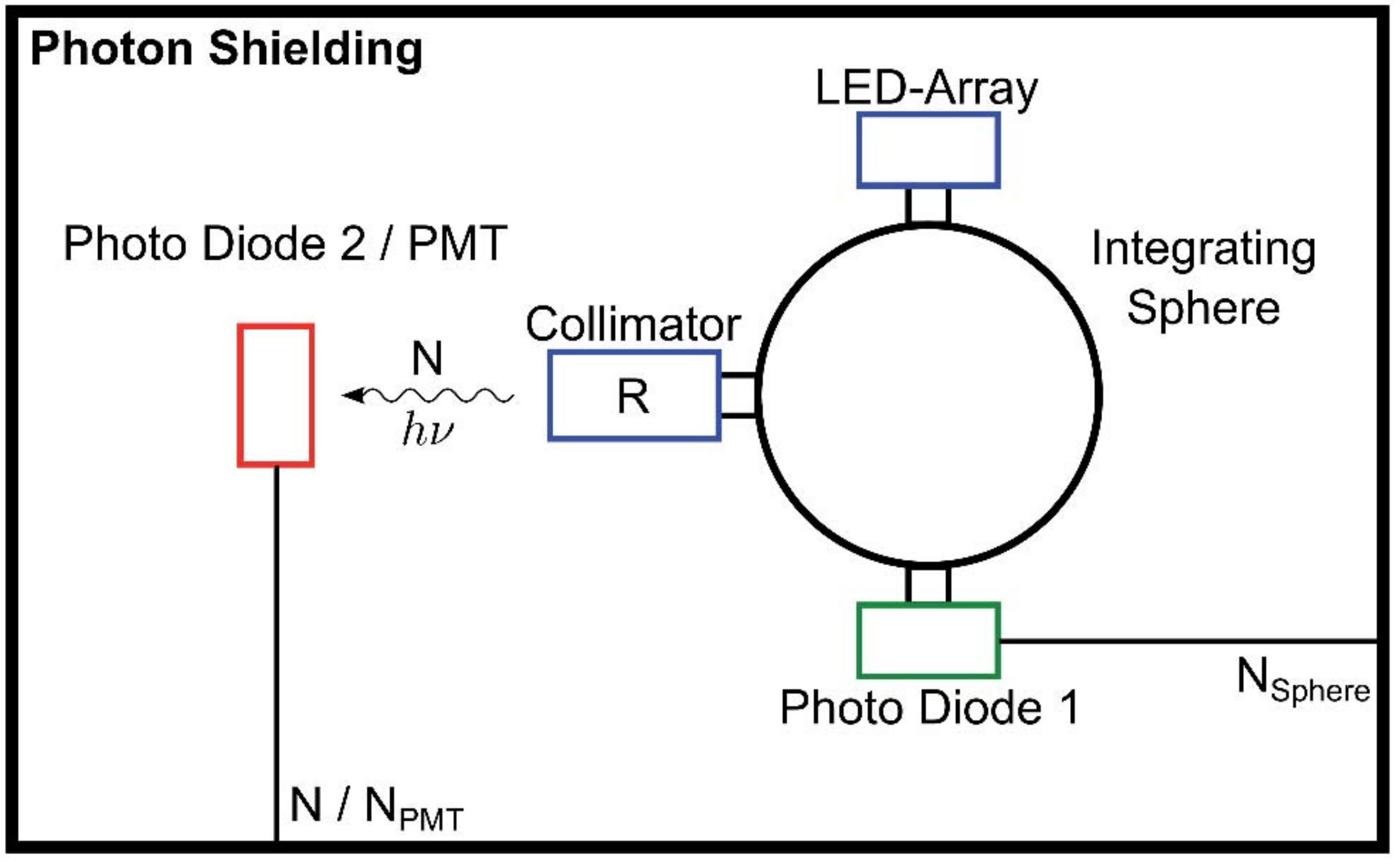}
\includegraphics[width=0.25\textwidth]{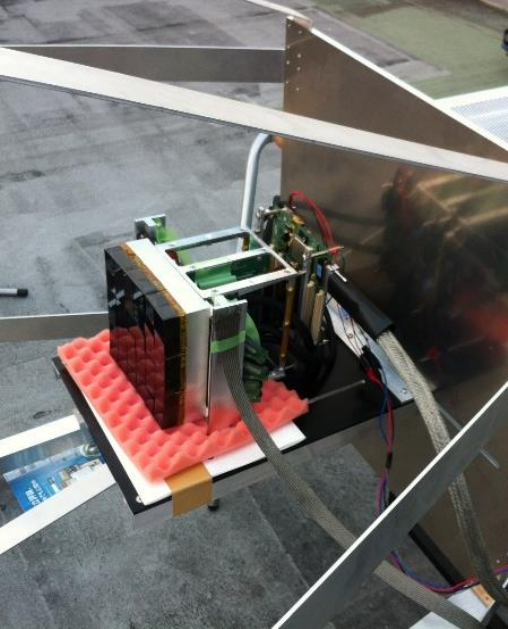}
\caption{Left: Sketch of the calibration stand for photo sensors. Right: first fully assembled Photo 
Detector Module (PDM) for JEM-EUSO prototype experiments.
\label{fig_calib}}
\end{figure}
The telescope (Fig.~\ref{fig_fs}) consists of four main parts: collecting optics with
3 Fresnel lenses, a focal surface detector, electronics and a structure.
JEM-EUSO constitutes a $\mu$s-fast and highly pixelized
digital camera with a diameter of about $2.5\,$m and a $\pm 30^\circ$ wide field of view. 
It works in the near-UV wavelength ($290-430\,$nm) and enables a single-photon 
counting. 

The focal surface will consist of 4932 64-channel multianode photomultipliers from Hamamatsu.
They will be organized in elementary cells of 4 PMs each and in 137 photo detection modules 
(PDM) of 9 elementary cells each. 
The focal surface (Fig.~ \ref{fig_fs}) comprises therefore more than 300,000 pixels in total. 
The electronics processes triggers for air-shower or other transient events in the 
Atmosphere and send necessary data to the ground for further analysis. 

The optics with three Fresnel lenses focuses the incident UV photons onto
the focal surface with an angular resolution of $0.07^\circ$. 

The Atmospheric Monitoring System (AMS)~\cite{EAams} consists of an infrared camera~\cite{EAir} and a
LIDAR (LIght Detection And Ranging) system, both monitoring the Earth's
atmosphere continuously inside the FoV of the JEM-EUSO telescope. 
A measurement of clouds top heights with an accuracy better than $500\,$m will be possible. 

\subsection{Calibration}

For fluorescence detection of cosmic rays it is essential to calibrate the detector pre-flight with utmost 
precision and to monitor the performance of the detector throughout the whole mission time.
For that purpose a calibration stand on-ground was built to measure precisely the performance of 
Hamamatsu 64\,pixel MAPMTs that are planned to be used for JEM-EUSO (Fig.~\ref{fig_calib}).
It consists of a photon shielding, a uniform light source and readout electronics for the MAPMTs. 
The uniform light is available thanks to an integrating sphere and the light intensity is monitored 
in real-time by a NIST-calibrated photo diode attached to the sphere.
To investigate the suitability of alternative detector devices, further research is done with 
state-of-the-art silicon photomultipliers (SiPMs). These will also be tested in the calibration stand and 
their performance can be compared to conventional photomultiplier tubes.
A designed inflight calibration system measures continuously the efficiencies of the optics, the focal
surface detector, and the data acquisition electronics~\cite{karus-icrc1,karus-icrc2}.

\subsection{Performance of JEM-EUSO}

\begin{figure}[ht]
\centering
\includegraphics[width=0.72\linewidth]{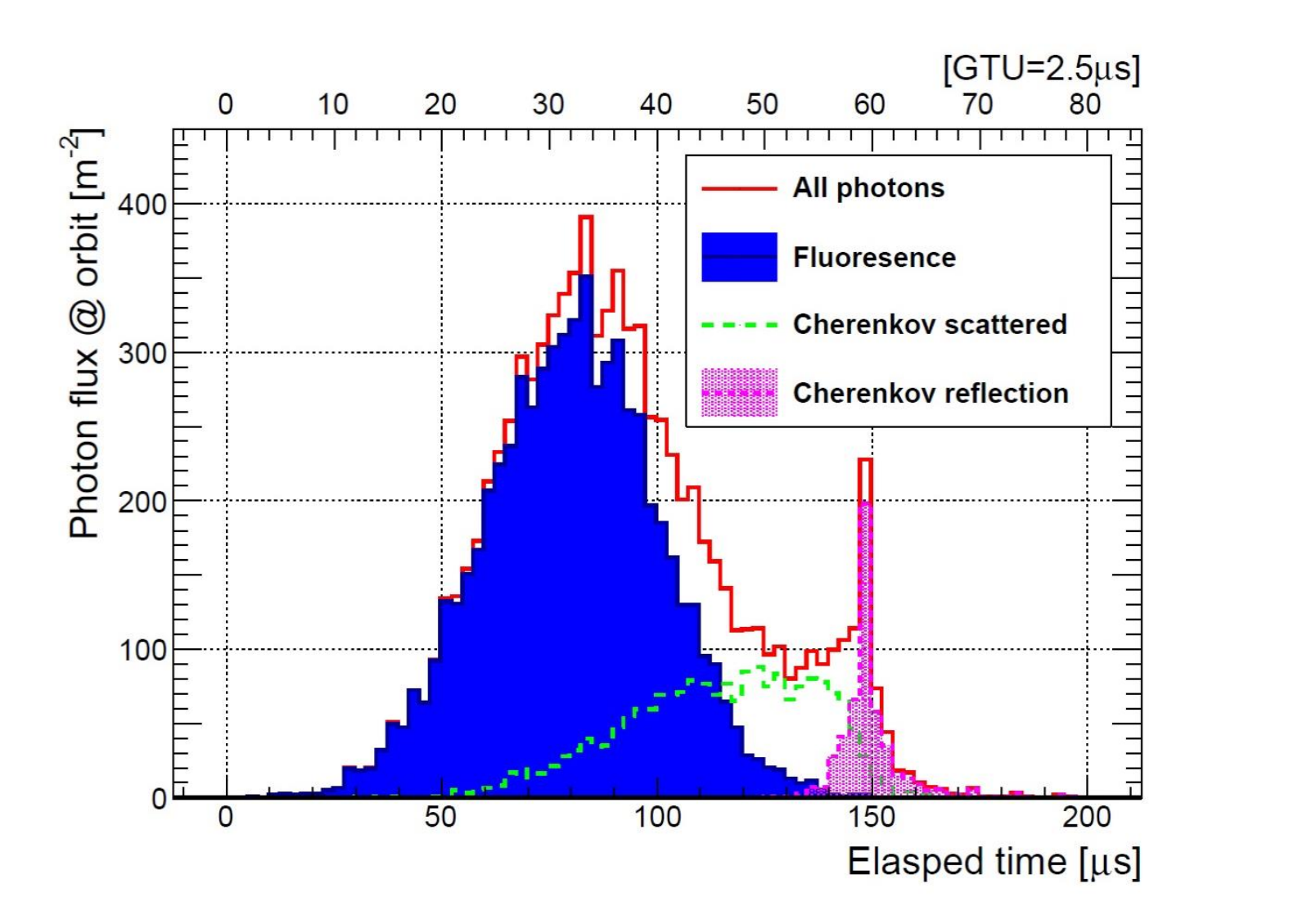}
\caption{Arrival time distribution of photons at the pupil of the telescope per
m$^2$ from a simulated high-energy EAS in case of 
a clear atmosphere.} 
\label{fig_signal}
\end{figure}
Figure~\ref{fig_signal} shows a typical longitudinal profile of an EAS at the pupil of the 
telescope. 
Simulations including the transmission losses of the optical system, the efficiency of 
the focal surface, as well as the rate of background photons show that JEM-EUSO 
reaches almost full efficiency already at energies around $3 \cdot 10^{19}\,$eV for
a restricted subset of events, and full aperture at energies $E > 5 - 6 \cdot 10^{19}\,$eV. 
The annual exposure of JEM-EUSO at $10^{20}\,$eV will exceed by one order of magnitude 
the level presently reached by ground-based observatories.
For clear atmospheric conditions a reconstruction of the shower geometry and primary energy 
with good accuracy, as well as a rough estimate of the shower maximum will be possible~\cite{fenu}.
In presence of optically thin clouds at high altitudes, often also categorized as cirrus, 
most of the EAS photons penetrate the layer of clouds and are attenuated only partly. 
This may lead to a reconstruction with a lower energy, however, the situation can be taken into 
account as the clouds are monitored by the AMS and the information is known for each triggered EAS. 
The geometry of the shower axis can still be properly determined by the analysis of the angular 
velocity of the EAS signal. 

\begin{figure}[t!]
\centering
\includegraphics[width=0.48\linewidth]{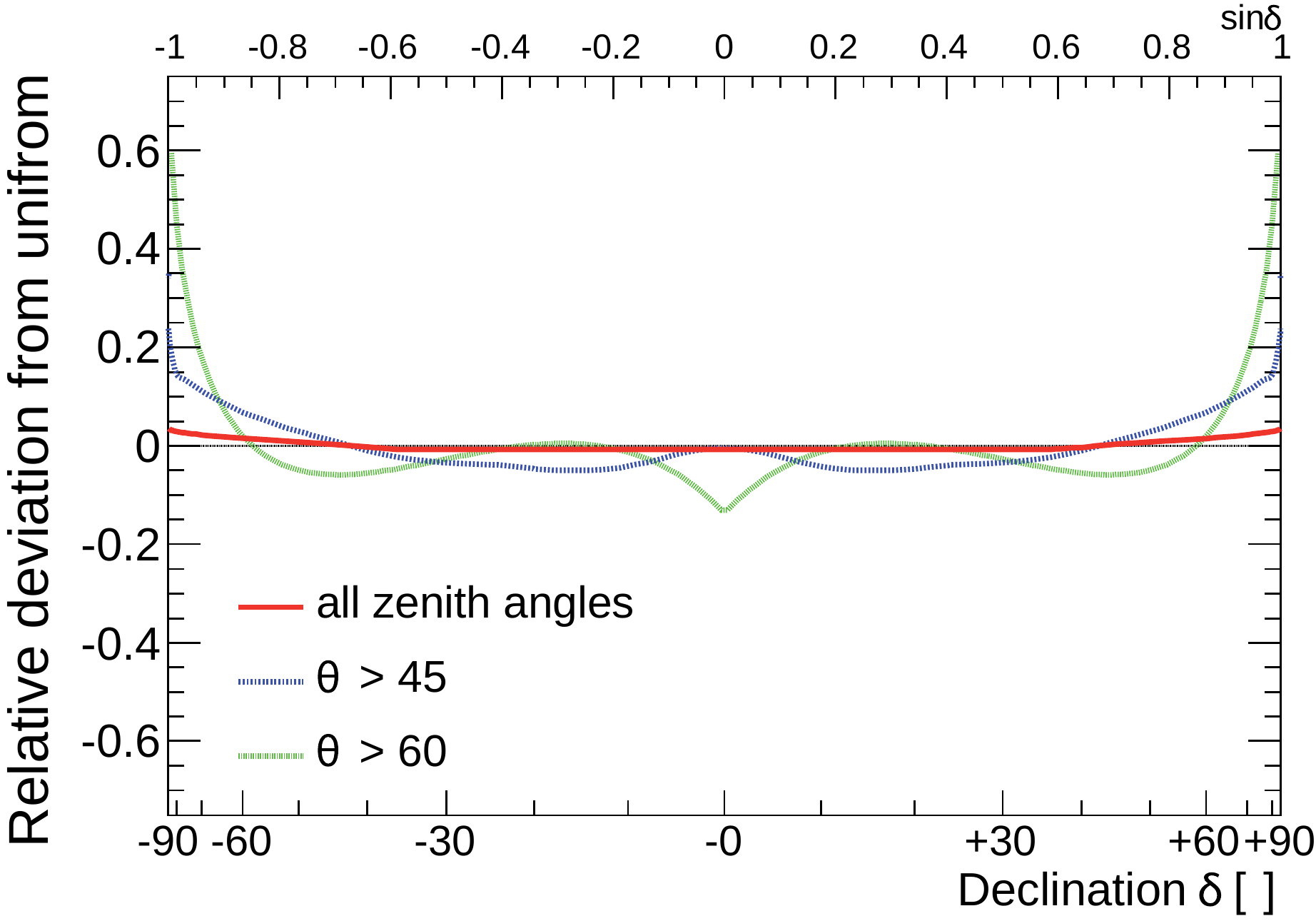}
\includegraphics[width=0.48\linewidth]{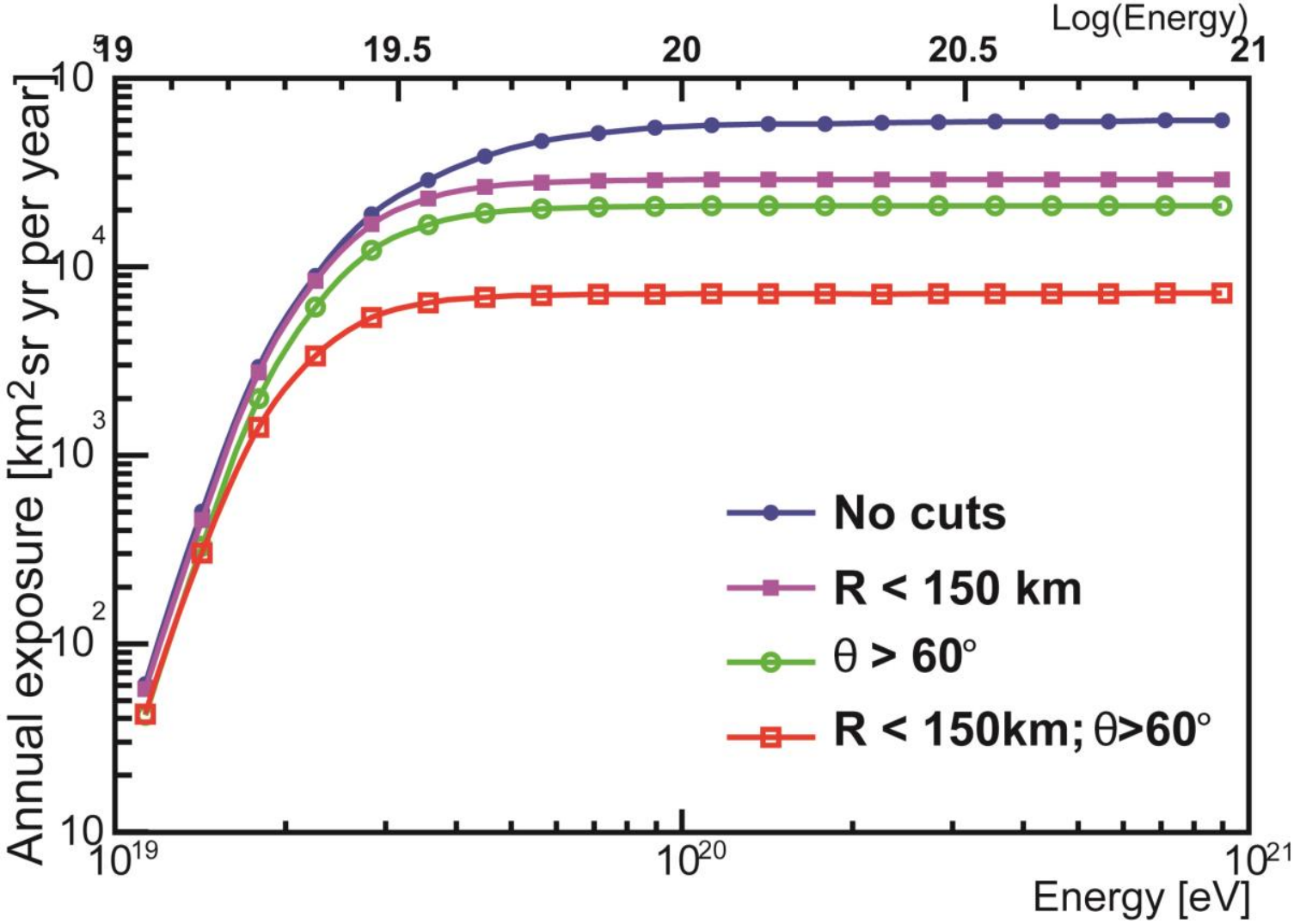}
\caption{Left: Relative deviation from uniformity of the aperture as a function
of sine of declination. The curves show the cases for a selection
of events with different zenith angle. 
Right: Annual exposure of JEM-EUSO (ISS at $400\,$km) for the
full FoV and for the case of extremely high quality cuts applied to the data. For the most 
stringent cuts, the threshold is low enough for an overlap with ground based experiments with good statistics.
Plots adapted from~\cite{perf-paper}.} 
\label{fig_perfor}
\end{figure}

One big advantage of JEM-EUSO will be the uniform full sky coverage due to the ISS orbit 
(Fig.~\ref{fig_perfor}, left). JEM-EUSO will be the first experiment to be able to take data of EECRs for 
both hemispheres to build a full sky map. This will lead to anisotropy studies and identification 
of sources and source regions of high-energy cosmic rays which is not possible with ground-based observatories.

The second advantage of the JEM-EUSO mission - compared to existing cosmic ray experiments using the 
fluorescence technique - is the big observation area on ground and therefore the big detector volume. 
This will enhance the statistics for UHECR events that are expected with a rate of one event per 
century per steradian per kilometer squared. 
JEM-EUSO will have a large observation area in nadir mode (studied here, Fig.~\ref{fig_perfor}, right panel) 
and even a roughly ten times bigger one in tilt mode (not yet fully explored, but the design enables 
the option to tilt the entire instrument at the ISS to have a larger coverage on ground).

JEM-EUSO triggers on the UV light from EAS, measures the intrinsic luminosity of each 
EAS near its maximum and accurately reconstructs the EAS path so that the arrival direction 
of the initiating EECR can be reconstructed. Detailed studies~\cite{perf-paper} indicate 
that for 68\% of the events the arrival direction can be reconstructed to better than
$2.5^\circ$.
Regarding the energy reconstruction, at the current
status of development of the
reconstruction algorithms, proton showers are reconstructed 
in clear-sky conditions with a typical energy resolution $\Delta E/E$ of 
$\approx$25\%(20\%) at energies around $4 \cdot 10^{19}\,$eV ($10^{20}\,$eV). 
In addition, our still preliminary results indicate
that the $X_{max}$ resolution can be better than $70\,$g/cm$^2$ for 
$E>10^{20}\,$eV in the central part of the FoV.
With 0.1 - 1 million km$^2$ sr exposure and the uniform coverage of the full sky, 
JEM-EUSO will observe all possible source directions at least within several hundred 
Mpc and will make possible the (i) identification of sources with unprecedentedly
high statistics by arrival direction analysis and (ii) the measurement of the energy 
spectra from individual sources to constrain the acceleration or the emission
mechanisms.

\section{Test Experiments}
\label{sec-test}

On the way to the full instrument ready for launch to the ISS, several test or pathfinder experiments 
are needed and currently under development or in operation.

A PDM (photon detection module with 2304 pixels, see Figs.~\ref{fig_fs} and \ref{fig_calib}) can be seen as an 
independent unit, where one of which are or will be used for most of the test experiments.  
However, new optical systems are adapted for the respective purpose.

\begin{itemize}
\item {\bf EUSO-Balloon:}
EUSO-Balloon~\cite{icrc-balloon} will serve as a demonstrator
for technologies and methods featured in the space instrument (Fig.~\ref{fig_TA}, right panel). 
This balloon-borne instrument points
towards the nadir from a float altitude of about $40\,$km.
With its Fresnel optics and PDM, the instrument monitors a 12 by 12 degree wide field of view. 
Key objectives are among others the full-scale end-to-end test of the JEM-EUSO technique, 
the experimental determination of the effective UV background below 40 km, 
ground based calibration systems, and, the observation of the first UV-image of an EAS looking down 
on the Earth's atmosphere.
The flights are handled by the balloon division of the French Space Agency CNES. 
A first flight was performed on August, 25, 2014 at Timmins, Canada. 5 hours 
of valuable data from the PDM, from the installed infrared camera, as well as Laser pulses 
sent by a helicopter accompanying the full flight are presently being analysed.

\item {\bf EUSO-TA:}
Also important information on the capabilities of the full instrument will be 
provided by EUSO-TA~\cite{icrc-ta}, 
which is a ground-based telescope formed by
one PDM (identical to the PDM of EUSO-Balloon) and two Fresnel lenses, which are prototypes of those
foreseen in JEM-EUSO. 
EUSO-TA will be located at the Telescope
Array (TA)~\cite{ta} site in Utah, USA (Fig.~\ref{fig_TA}, left panel). 
The instrument will measure the UV light in the Atmosphere
in its $\pm4$ degrees FoV, will be triggered by the fluorescence detectors of TA, 
and will make use of the LIDAR and the Electron Light Source of TA.  
The aims are to obtain an end-to-end calibration of the prototype telescope, and
an inter-calibration with the fluorescence detector of TA.
Everything is prepared to start the measurements in early 2015.  
\begin{figure}[t!]
\centering
\includegraphics[width=0.52\linewidth]{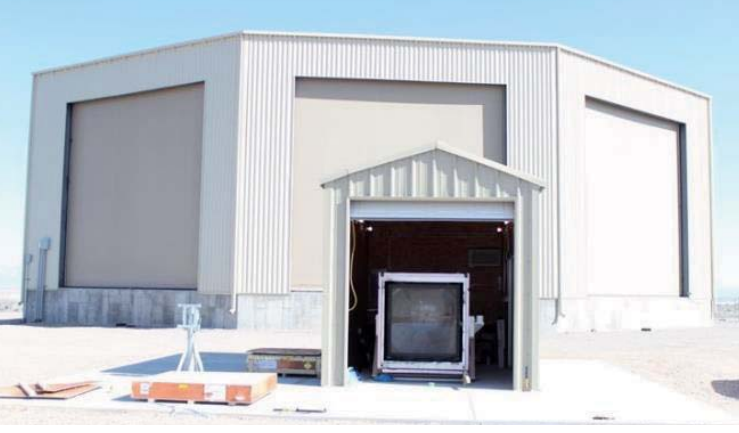}
\includegraphics[width=0.165\linewidth]{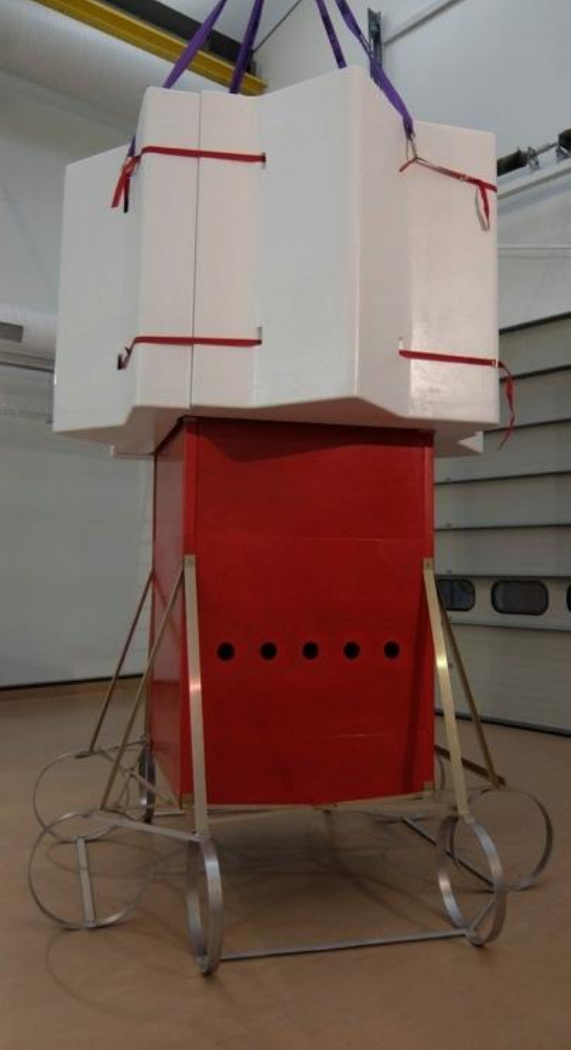}
\caption{Left: photo of the EUSO-TA installation in the front of the Telescope Array fluorescence 
detector in Utah, US. Right: photo of the EUSO-Balloon payload (with floating devices) just before 
the launch in August 2014 in Timmins, Canada.} 
\label{fig_TA}
\end{figure}

\item {\bf K-EUSO:}
Currently, the collaboration considers a cooperation with the Russian project KLYPVE. KLYPVE~\cite{klypve} 
is a mirror based fluorescence telescope foreseen to be installed at the Russian segment of the ISS. 
Despite its smaller size, KLYPVE could be the first experiment detecting EAS from space. 
Combined with a Fresnel lens and a JEM-EUSO type focal surface the exposure and quality of the 
experiment could be considerably improved. However, due to limitations in space and weight, 
the exposure of the baseline JEM-EUSO design probably can not be reached.    

\item {\bf Mini-EUSO:}
Another project is Mini-EUSO, a small prototype experiment (one PDM and a two-25cm-diameter Fresnel-lens 
system) foreseen to operate inside the ISS for an observation of the UV emission 
from the night-Earth through an UV-window. 
This instrument would map for the first time the Earth in UV, and could 
study atmospheric phenomena and bio-luminescence at Earth as well as meteors.  

\item {\bf SiPM-EUSO:}
The collaboration has also started to evaluate if SiPM can replace the heavy and 
expensive Multianode PMT's presently in use.
Designing and manufacturing an R\&D early prototype of a SiPM based PDM is under study for the 
use in the test experiments. 
\end{itemize}

\begin{figure}[ht]
\centering
\includegraphics[width=0.66\linewidth]{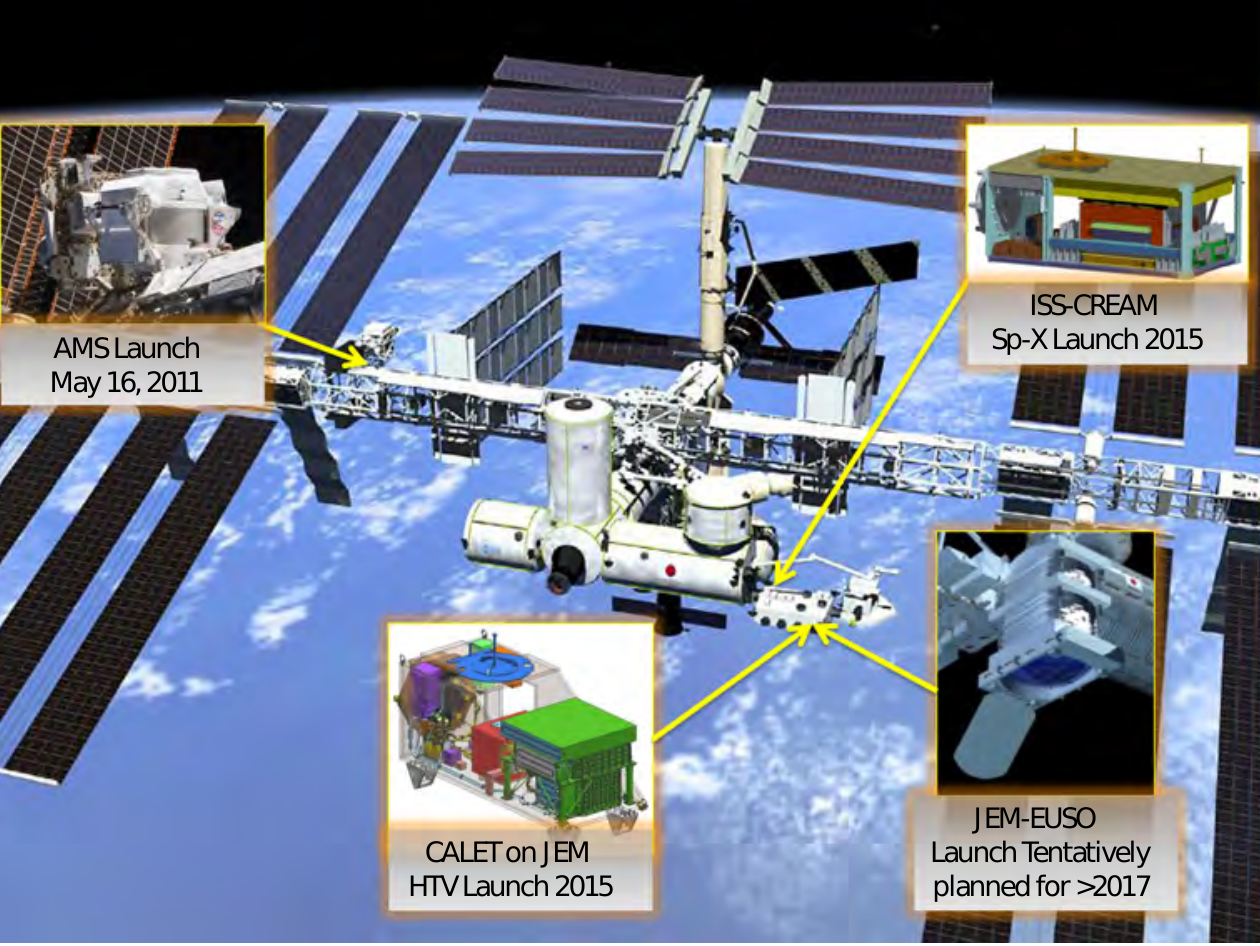}
\caption{Illustration of four cosmic ray instruments on the ISS, which would comprise complementary 
studies of a wide field in cosmic ray physics. Picture taken from~\cite{jones}.}  
\label{fig_nasa}
\end{figure}

\section{Outlook}

JEM-EUSO is planned as a three to five year mission, where initially the launch was foreseen in 2017. 
Due to financial reasons this will not be possible. However, the JEM-EUSO collaboration continues the 
efforts to improve the baseline design and the capabilities and sensitivity of such an instrument with the
aim to launch it to space at a later opportunity.   
Still the idea lives on of a comprehensive, virtual wide-range Cosmic Ray Observatory in space, 
like outlined by Vernon Jones from NASA, e.g.~(Figure~\ref{fig_nasa}).  

The various test and pathfinder experiments presently under construction or in operation will provide 
more information and details on the capabilities of a space instrument like JEM-EUSO.

\ack
Supported by the `Helmholtz Alliance for Astroparticle Physics HAP' funded by the 
Initiative and Networking Fund of the Helmholtz Association.

\section*{References}

\end{document}